\documentclass[aps,prl,twocolumn,showpacs,superscriptaddress,floatfix]{revtex4-1}

\usepackage{amsmath,amssymb}
\usepackage{graphicx}
\usepackage{color}
\usepackage{setspace}
\usepackage{hyperref}
\usepackage{bm}

\newcommand{\PrCa}{Pr$_{x}$Ca$_{1-x}$Fe$_2$As$_2$}
\newcommand{\dPrCa}{Pr$_{0.105}$Ca$_{0.895}$Fe$_2$As$_2$}

\begin{document}

%\preprint{}

\title{Nanoscale Surface Element Identification and Dopant Homogeneity\\in the High-$T_c$ Superconductor Pr$_x$Ca$_{1-x}$Fe$_2$As$_2$}

\author{Ilija Zeljkovic}
\author{Dennis Huang}
\author{Can-Li Song}
\affiliation{Department of Physics, Harvard University, Cambridge, MA 02138, U.S.A.}
\author{Bing Lv}
\author{Ching-Wu Chu}
\affiliation{Texas Center for Superconductivity, University of Houston, Houston, TX 77204, U.S.A.}
\author{Jennifer E Hoffman}
\email[]{jhoffman@physics.harvard.edu}
\affiliation{Department of Physics, Harvard University, Cambridge, MA 02138, U.S.A.}
\date{\today}


\begin{abstract}

We use scanning tunneling microscopy to determine the surface structure and dopant distribution in \PrCa, the highest-$T_c$ member of the 122 family of iron-based superconductors. We identify the cleaved surface termination by mapping the local tunneling barrier height, related to the work function. We image the individual Pr dopants responsible for superconductivity, and show that they do not cluster, but in fact repel each other at short length scales. We therefore suggest that the low volume fraction high-$T_c$ superconducting phase is unlikely to originate from Pr inhomogeneity.

\end{abstract}

% 68.37.Ef  Scanning tunneling microscopy (including chemistry induced with STM)
% 74.55.+v  Tunneling phenomena: single particle tunneling and STM
% 74.70.Xa  Pnictides and chalcogenides
% 74.62.Dh  Effects of crystal defects, doping and substitution


\maketitle
The recent discovery of high-$T_c$ superconductivity in Fe-based materials\cite{Kamihara2008iron} has rejuvenated worldwide efforts to understand and predict new superconductors. Like cuprates, Fe-based superconductors (Fe-SCs) are layered, with Fe-based superconducting planes separated by buffer layers. Furthermore, superconductivity typically arises by chemically doping an antiferromagnetic parent compound \cite{Johnston2010puzzle}. In the first generation of $A$Fe$_2$As$_2$ (122) Fe-SCs, hole doping resulted in higher maximum $T_c$ ($38\;\mathrm{K}$ in K$_x$Ba$_{1-x}$Fe$_2$As$_2$ \cite{Rotter2008superconductivity}) than electron doping ($25\;\mathrm{K}$ in Ba(Fe$_{1-x}$Co$_x$)$_2$As$_2$ \cite{Nakajima2009possible}). However, the highest $T_c$ among all Fe-SCs was $57\;\mathrm{K}$ in electron-doped Sm$_{1-x}$La$_x$O$_{1-y}$F$_y$FeAs \cite{Wei2008superconductivity}, prompting the suggestion that $T_c$ could be enhanced in electron-doped 122s by removing the damaging dopant disorder from the crucial Fe layer, and doping the buffer layer instead. The strategy was successful in the rare-earth-doped Ca122 family \cite{Saha2012structural,Lv2011unusual,Qi2012Ca122}, with $T_c$ reaching $49\;\mathrm{K}$ in \PrCa. However, the high $T_c$ appeared in only $\sim 10\%$ of the volume, while the bulk of the material showed $T_c \sim 10-20\;\mathrm{K}$.

% Note the Wei 2008 paper is kind of sketchy because it doesn't zoom in on the transition.

Saha \textit{et al.} performed a thorough search for the origin of the low volume fraction high-$T_c$ phase, using bulk experimental probes.  First, the high $T_c$ was found to be impervious to etching or oxidation, arguing against surface superconductivity. Second, high-$T_c$ resistive transitions were never observed for dopant concentrations below those necessary to suppress the parent antiferromagnetic phase, arguing against random inclusions as the origin. Furthermore, no such contaminant phases were observed in over 20 samples examined by x-ray diffraction. Third, the high $T_c$ was unaffected by the global structural collapse phase transition (the abrupt $\sim10\%$ shrinkage of the $c$-axis lattice constant that occurs in the Ca122 family under external or chemical pressure), arguing against any relationship to the collapsed phase or to interfaces between collapsed and non-collapsed phases. In fact, aliovalently-doped CaFe$_2$(As$_{1-x}$P$_x$)$_2$ also shows the structural collapse but no high-$T_c$ volume fraction \cite{Kasahara2011abrupt}. Saha \textit{et al.} therefore concluded that the charge doping is an essential ingredient to the high-$T_c$ phase, and speculated that it has ``a localized nature tied to the low percentage of rare earth substitution.''

Given the challenges in identifying the origin of the low volume fraction high-$T_c$ phase from bulk experiments, a local probe is naturally required. Here we use scanning tunneling microscopy (STM) to investigate two possible sources of electronic inhomogeneity in \PrCa: the surface and the dopants. We provide the first definitive identification of the cleaved surface termination and the first image of all individual dopants in the Ca122 system. Based on our results, we suggest that dopant inhomogeneity is unlikely to be responsible for the low volume fraction of high-$T_c$ superconductivity in \PrCa.

Single crystals of \PrCa\ are grown via self-flux with measured $x=10.5\%$ and resistive $T_c=43.2\;\mathrm{K}$ (See Supplementary Information \textbf{S-I}). The crystals are handled exclusively in Ar environment, cleaved in ultra-high vacuum at cryogenic temperature, and immediately inserted into the STM head where they are imaged with a PtIr tip, cleaned by field emission on Au. The first challenge in STM imaging of any new material is to identify the surface structure and evaluate to what extent it is representative of the bulk. The surface structure of the $A$Fe$_2$As$_2$ system has been particularly controversial \cite{Hoffman2011spectroscopic}. Due to the stronger bonding within the FeAs layer [Fig.\ 1(a)], the FeAs layer is expected to remain intact upon cleaving, leaving half a complete $A$ layer on each surface \cite{Gao2010surface}. This preservation of charge neutrality is a necessary (but not sufficient) condition for the surface to be representative of the bulk.  However, a number of experiments have claimed that the cleaved surface is As-terminated in the Ba122 \cite{Nascimento2009surface}, Sr122 \cite{Niestemski2009unveiling}, and Ca122 \cite{Huang2012experimental} systems.

We encounter three different surface morphologies in our STM topographs of \PrCa. The majority of the observed sample surface displays a $2\times 1$ structure [Fig.\ 1(b)] frequently observed in other STM studies of 122 materials \cite{Massee2009cleavage}. We occasionally observe a disordered, ``web-like'' structure [Fig.\ 1(c)], which smoothly merges with the $2\times 1$ structure [Fig.\ 1(d)]. The third type of surface, observed rarely, shows a $1\times 1$ square lattice with $\sim4\;\mathrm{\AA}$ periodicity [Fig.\ 1(e)].

\begin{figure}[tb]
\begin{center}
\includegraphics[width=0.99\columnwidth,clip]{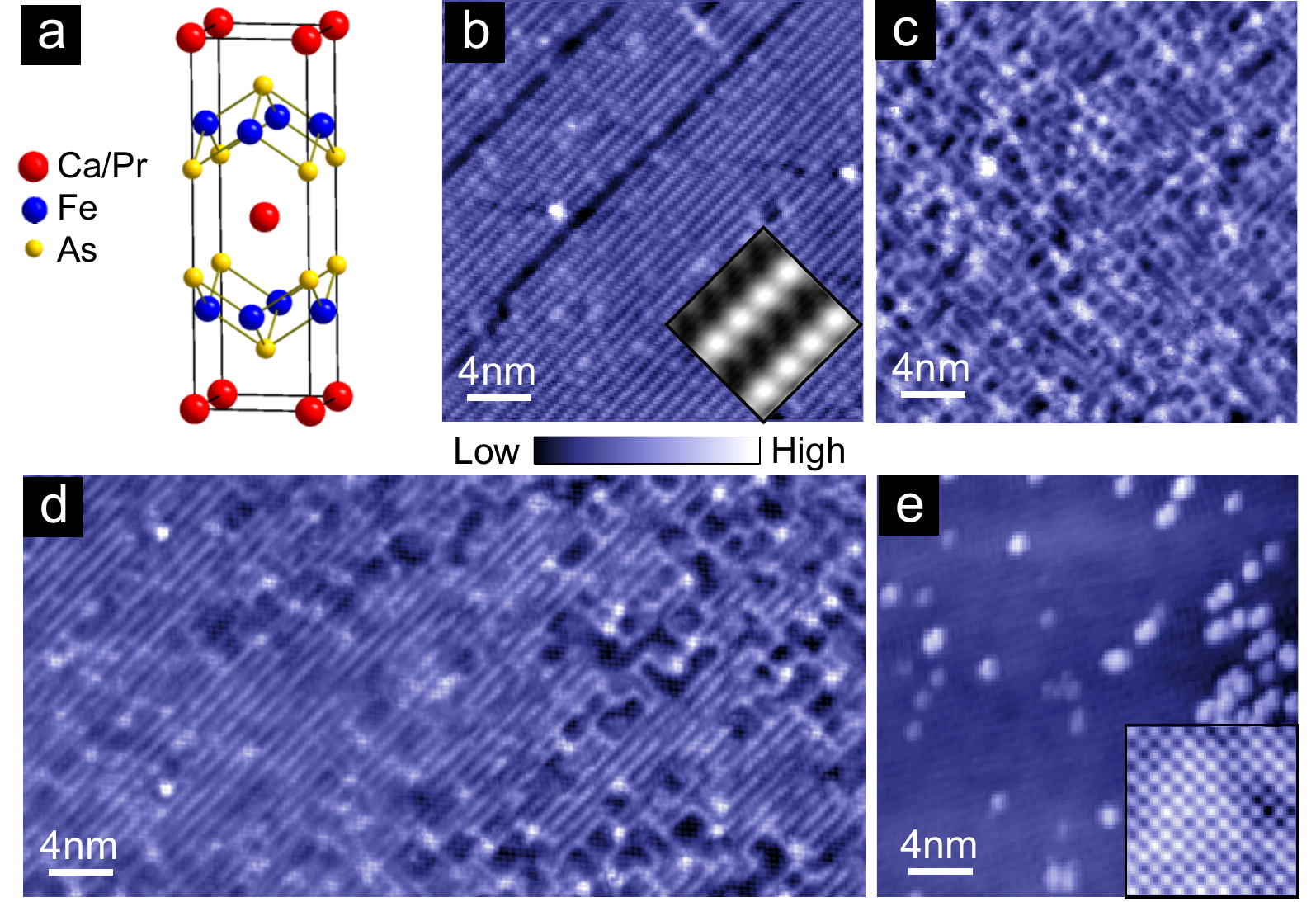}
\caption{(color online) Surface morphologies of cold-cleaved \dPrCa. (a) Crystal structure of \PrCa. Topographs of (b) $2\times 1$ surface structure (250 pA, +300 mV, 7 K) (c) disordered, ``web-like'' surface structure (15 pA, 100 mV, 7 K), (d) smooth transition between $2\times 1$ and ``web-like'' structures (20pA, -100mV, 7K), and (e) $1\times 1$ square lattice with $\sim4\;\mathrm{\AA}$ lattice constant (5 pA, 50 mV, 25 K). Inset in (b) shows an average $30\times 15$ pixel, $2\times 1$ supercell \cite{Zeljkovic2012scanning} tiled $2 \times 4$ times. Inset in (e) shows an enlarged, 4 nm topograph of $1\times 1$ square lattice acquired at 50 mV and 50 pA. (Due to an external noise source present during the acquisition of the data in panel (e), the images in (e) have been filtered to remove all spurious spatial frequencies higher than the $4\;\mathrm{\AA}$ periodicity.)
}
\end{center}
\end{figure}

% Former fig 1(b) $2\times 1$ reconstruction (20 pA, 100 mV, 7K)
% New fig 1(b) is the old fig 3(a), 25 nm FOV

\begingroup
\begin{table}
\caption{\label{tab:workfunction}Work functions for several pure elements. \cite{Speight2005handbook}}
\begin{tabular}{ l l l l l l l l l l}
 \hline
    Atom & Fe & As & Ca & Pr & Sr & Ba & Au & Pt & Ir \\
    $\varphi$ (eV) & 4.65 & 3.75 & 2.71 & 2.7 & 2.76 & 2.35 & 5.32 & 5.40 & 5.6\\
  \hline
\end{tabular}
\end{table}
\endgroup

We map the tunneling barrier height to identify these surfaces. The tunneling current $I$ is expected to decay exponentially with the tip-sample separation $z$ as
\[
%\begin{equation}
%\label{eq:tunneldecay}
I \propto e^{-\sqrt{\frac{8 m_e \Phi}{\hbar^2}} z} \nonumber
%\end{equation}
\]
\noindent where $\Phi$ is the local barrier height (LBH), approximately equal to the average of the tip and sample work functions \cite{Binnig1984electron}. However, the LBH is sensitive not only to the elemental composition of the tip, but also the geometric configuration of the tip's terminal atoms, and the tip-sample angle (which reduces the LBH by $\cos^2 \theta$, where $\theta$ is the deviation between the sample surface perpendicular and the $z$ direction of tip piezo motion \cite{Wiesendanger1987local}). Moreover, the LBH depends on the sample topography through two opposing mechanisms. On the one hand, protruding atoms or clusters may stretch out as the tip is retracted, reducing the effective rate at which the tip-sample distance decreases, and thus suppressing the measured LBH above the protrusion \cite{Yamada2003local}. On the other hand, the topographic corrugation appears smoothed out at distances far from the surface; this implies that the wave function decays faster above a protrusion than a depression, thus enhancing the measured LBH above a protrusion \cite{Wiesendanger1994scanning}. Without accounting for these factors, previous studies found the LBH on the $2\times 1$ surface of BaFe$_2$As$_2$ to be much lower than the expected work functions for either Ba or As \cite{Massee2011tunnelers}.

In contrast, the comparison of LBH measurements with the same tip (i.e.\ the same microscopic configuration of terminating atoms) across different flat regions of the same cleaved surface (i.e.\ the same tip-sample angle) can yield a robust measure of relative work functions, and can be utilized for element identification in cases where the sample consists of two different surfaces \cite{Jia1998variation}. Here, we directly compare LBH values measured with the same STM tip across the different morphologies of Fig.\ 1 on the same cleaved sample.

To extract the LBH at each point $(x,y)$ in a field of view (FOV), a feedback loop first adjusts $z_0(x,y)$ to maintain $I=100\;\mathrm{pA}$ at $V_{\mathrm{set}}=-100\;\mathrm{mV}$; the current $I(z)$ is then measured as the tip is retracted from $z_0$. Figure 2 shows simultaneous topographs and work function maps for the $1\times 1$ square lattice across a step edge [Figs.\ 2(a,c)] and the $2\times 1$ structure in a nearby flat area [Figs.\ 2(b,d)]. Figure 2(e) shows two sets of representative $I(z)$ curves from the square regions in Figs.\ 2(a,b), which are clearly distinct from one another. After correcting for the surface slope,
% The x and y angles were 6.8 and 17.7 degrees. The combined angle was 18.9 degrees.
we find the average LBH values are $\Phi_{1\times 1} = 4.50 \pm 0.42\;\mathrm{eV}$ and $\Phi_{2\times 1} = 3.57 \pm 0.34\;\mathrm{eV}$. Taking Table~\ref{tab:workfunction} and the tip work function into account, these values suggest that the $1\times 1$ surface is a complete As layer, while the $2\times 1$ surface is a half Ca layer.

\begin{figure}[tb]
\begin{center}
\includegraphics[width=0.85\columnwidth,clip]{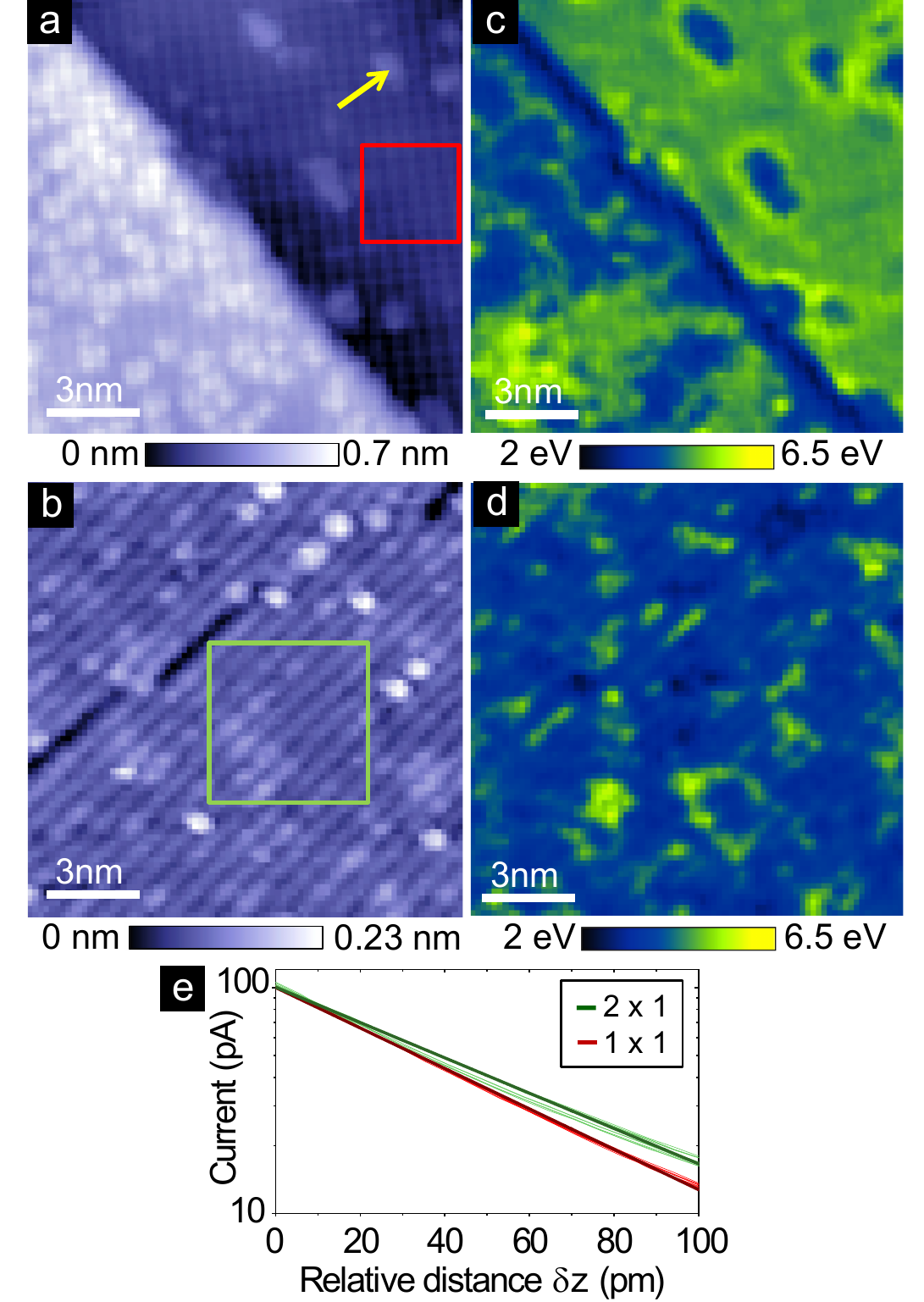}
\caption{(color online) LBH comparison between $1\times 1$ and $2\times 1$ surface structures. Topographs acquired at 25 K of (a) $1\times 1$ (4 \AA\ $\times$ 4 \AA) square lattice appearing on both sides of a step edge and (b) $2\times 1$ (8 \AA\ $\times$ 4 \AA) lattice. Simultaneously acquired LBH maps are shown in (c) and (d). Approximately 26\% larger LBH is observed in a clean, flat area of the $1\times 1$ surface (red box in (a)) than in a clean flat area of the $2\times 1$ surface (green box in (b)). Furthermore, sparse topographic protrusions on the $1\times 1$ surface (e.g.\ marked by yellow arrow) show a lower LBH close to that of the flat $2\times 1$ surface, suggesting that they are scattered remaining Ca or Pr atoms (See Supplementary Information \textbf{S-II}). Both datasets were acquired at $I_{\mathrm{set}}$=100 pA and $V_{\mathrm{set}}$=-100 mV. (e) Representative sets of $I(z)$ curves from square regions in (a) and (b) are shown as thin red and green lines respectively. Darker red and green lines represent linear fits to the average $I(z)$ curves from boxes in (a) and (b).}
\end{center}
\end{figure}

%Furthermore, the LBH measured on top of bright atoms observed in Fig.\ 2(a) is approximately equal to $\Phi_{2\times 1}$, consistent with those atoms being leftover Ca/Pr atoms on top of the As surface.

We expect the LBH to be affected by differences in the local environments for As/Ca atoms on \PrCa\ versus their respective pure elements. A dominant contribution to these differences may be due to dipole barriers arising from charge redistribution at the surface \cite{Bardeen1935theory}. This can increase or decrease the measured LBH according to the relative sign and magnitude of surface dipole barrier on \PrCa\ versus the pure element single crystals represented in Table I. We therefore expect
\[
%\begin{equation}
%\label{eq:LBHdipole}
\Phi_{1\times 1} = \frac{\varphi_{\mathrm{As}} + E_{\mathrm{As}}^{\mathrm{dip}} + \varphi_{\mathrm{tip}}}{2};\; \Phi_{2\times 1} =
\frac{\varphi_{\mathrm{Ca}} + E_{\mathrm{Ca}}^{\mathrm{dip}} + \varphi_{\mathrm{tip}}}{2} \nonumber
%\end{equation}
\]
\noindent where $\varphi_{\mathrm{As}}$ and $\varphi_{\mathrm{Ca}}$ are the pure element work functions, and $E_{\mathrm{As}}^{\mathrm{dip}}$ and $E_{\mathrm{Ca}}^{\mathrm{dip}}$ are the additional energies for an electron to escape the dipole layers at the As- and Ca-terminated surfaces of \PrCa. Assuming $\varphi_{\mathrm{tip}} \lesssim \varphi_{\mathrm{PtIr}}$ \cite{tip}, the values of $E_{\mathrm{As}}^{\mathrm{dip}}$, $E_{\mathrm{Ca}}^{\mathrm{dip}}$, and their difference $E_{\mathrm{As}}^{\mathrm{dip}} - E_{\mathrm{Ca}}^{\mathrm{dip}} = 0.82\;\mathrm{eV}$ are all of the correct magnitude for such dipole layers \cite{Bardeen1935theory}. The sign of the difference, which indicates that it is harder to remove an electron from the dipole barrier of the As surface than that of the half-Ca surface, is physically justified because the half-Ca surface is nonpolar, whereas the As surface is deficient of electrons from the stripped Ca, and thus more electronegative. The inferred electron-deficiency of this As surface is consistent with the failure to observe even proximity-induced superconductivity on the As-terminated surface of the related Sr$_{0.75}$K$_{0.25}$Fe$_2$As$_2$ \cite{Song2012electronic}.

The density of atoms in the surface layer is also known to affect the measured LBH \cite{Lang1971MetalLBHtheory,Gartland1972Cu}. Theories that treat the ionic lattice pseudopotential as a perturbation predict $\sim$10\% variations among different faces of single crystals \cite{Lang1971MetalLBHtheory}. Measurements on Cu(100), Cu(110), Cu(112), and Cu(111) confirm variations of at most 10\% \cite{Gartland1972Cu} although the surface density of atoms differs by almost 65\% between Cu(110) and Cu(111). Therefore, the 50\% difference in atom density between 1 x 2 and 1 x 1 surfaces of \PrCa\ is unlikely to account for the 26\% difference we observe between LBH values on these surfaces (Fig.\ 2).

% $\Phi_{1\times 1} > \Phi_{2\times 1}$ by 26\%, while for the elemental work functions, $\varphi_{\mathrm{As}} > \varphi_{\mathrm{Ca}}$ by 38\%

% Note that we refer to the "2x1 surface", not the "2x1 reconstruction" because the 2x1 is a half layer, i.e. it's actually 2x1 atoms, rather than being a reconstruction of a complete layer.

We also note the reduced LBH along the step edge in Fig.\ 2(c), which may be attributed to two mechanisms \cite{Jia1998variation}. First, the step edge is effectively an angled surface, so the LBH is reduced by $\cos^2 \theta$. Second, the Smoluchowski smoothing of the electron wave functions along the step edge results in an additional dipole moment which reduces the LBH \cite{Smoluchowski1941anisotropy}.

% Note that Wiesendanger 1987 does NOT see LBH reduction along the graphite step edge.  (The line of reduced work function in his image is not a step edge; it's some other unidentified linear defect.  Whereas the step edge on the left side shows barely any reduced work function.)

\begin{figure}[tb]
\begin{center}
\includegraphics[width=0.85\columnwidth,clip]{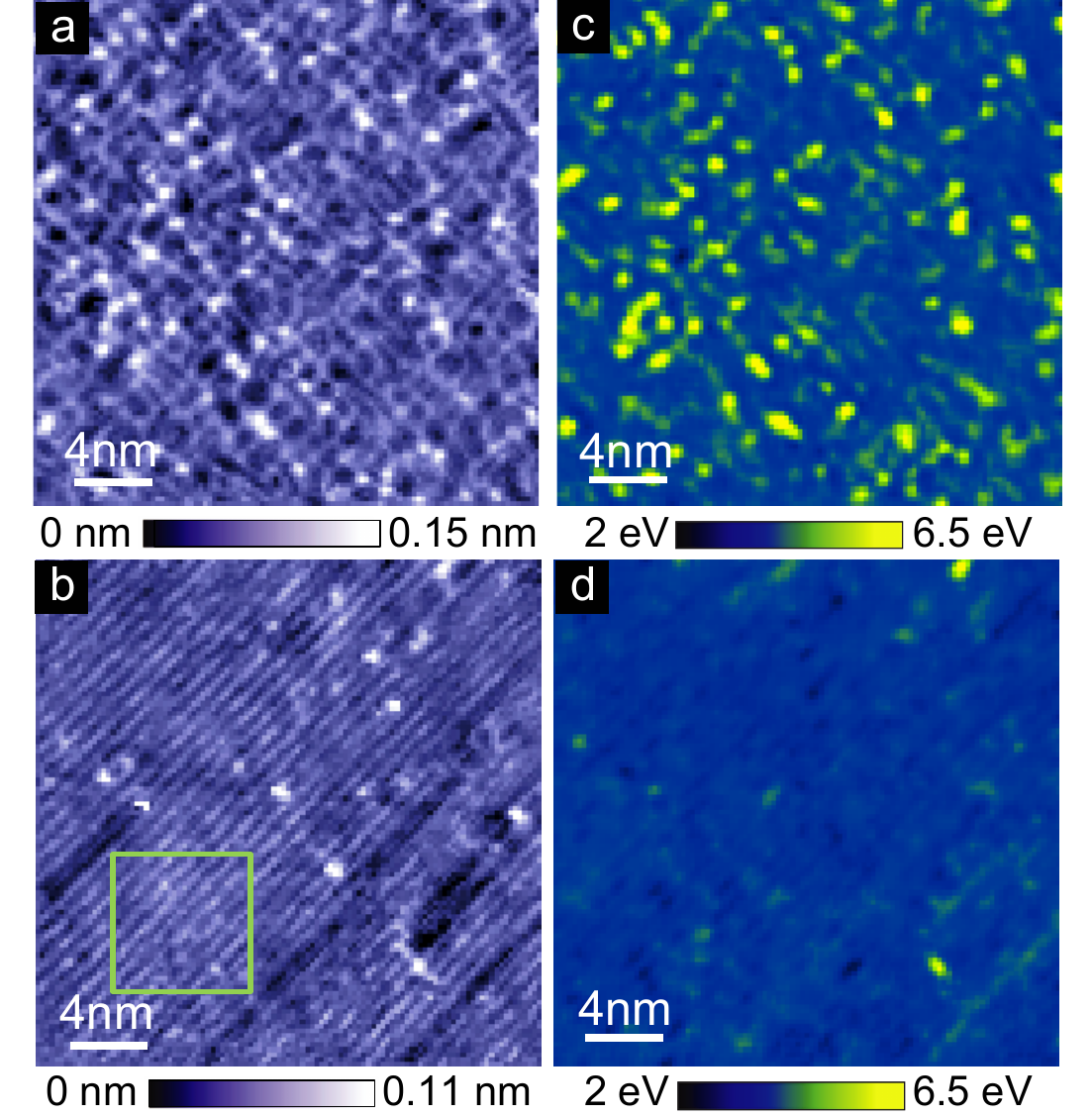}
\caption{(color online) LBH comparison between ``web-like'' and $2\times 1$ surface structures. Topographs acquired at 7 K of (a) ``web-like'' surface and (b) $2\times 1$ surface. Simultaneously acquired LBH maps are shown in (c) and (d). Both datasets were acquired at $I_{\mathrm{set}}$=105 pA and $V_{\mathrm{set}}$=100 mV. The $z$ calibration used here was obtained by assuming that the average LBH for the $2\times 1$ surface in the boxed region of (b) here is the same as that in Fig.\ 2(d).
}
\end{center}
\end{figure}

To further support the identification of the $2\times 1$ surface, we show a high-resolution map of the intra-unit cell structure to rule out the possibility of ``hidden'' surface atoms. We correct for small piezoelectric and thermal drift by placing the Ca/Pr atoms of Fig.\ 1(b) on a perfect lattice \cite{Lawler2010intra}. We then use the whole FOV to create the average $2\times 1$ supercell in the inset to Fig.\ 1(b) \cite{Zeljkovic2012scanning}. We do not observe atom dimerization (as seen in Ca$_{0.83}$La$_{0.17}$Fe$_2$As$_2$ \cite{Huang2012experimental} and Sr$_{1-x}$K$_x$Fe$_2$As$_2$ \cite{Niestemski2009unveiling,Boyer2008scanning}), but rather a single row of atoms, similar to the CaFe$_2$As$_2$ parent compound \cite{Chuang2010nematic}. We note that the appearance of dimerization \cite{Niestemski2009unveiling,Huang2012experimental,Boyer2008scanning} may be an artifact attributed to hybridization between the surface $A$ atoms and the underlying As atoms \cite{Gao2010surface}.

For completeness, we investigate the nature of the ``web-like'' surface. Because it merges smoothly into the $2\times 1$ surface without any evident step edges [Fig.\ 1(d)], it is also likely a reconstruction of the Ca layer. A simultaneous topograph and LBH map of the ``web-like'' surface are shown in Figs.\ 3(a,c), with analogous maps for the $2\times 1$ surface, acquired with the same tip for direct comparison, shown in Figs.\ 3(b,d). Bright spots in the topograph of Fig.\ 3(a) exhibit anomalously high LBH, highlighting the importance of the complex geometric effects of protrusions previously mentioned (See Supplementary Information \textbf{S-II}).  This reinforces the necessity of flat atomic planes in order to extract a reliable LBH comparison. We reiterate that our identification of the $1\times 1$ surface as a complete As layer and the $2\times 1$ surface as a half-Ca layer is robustly drawn from the flat surfaces in Fig.\ 2.

Since $\varphi_{\mathrm{Ca}}$ and $\varphi_{\mathrm{Pr}}$ differ by less than 1\% (Table~\ref{tab:workfunction}), LBH mapping cannot be used to identify Pr atoms in the Ca surface layer. However, STM can image dopants using the differential conductance $dI/dV$, which is proportional to the local density of states \cite{Zeljkovic2012imaging}. Substituting Pr$^{3+}$ for Ca$^{2+}$ creates a localized positive charge, so the impurity state is expected above the Fermi level.  We therefore search for Pr dopants in $dI/dV$ images at high bias. Figure 4(a) shows a $dI/dV$ image obtained simultaneously with the topograph in Fig.\ 1(b) at $+300\;\mathrm{mV}$, revealing a set of bright, atomic-scale features that can be visually identified in the simultaneous topograph with constituent atoms of the $2\times 1$ surface. These features, which start to appear in $dI/dV$ at biases higher than $+70\;\mathrm{mV}$, comprise $\sim$10.4\% of the total number of visible atoms in this FOV, matching the macroscopically measured $x=10.5\%$ and confirming the half-Ca termination. Although a subset of Co dopants were previously imaged in Ca(Fe$_{1-x}$Co$_x$)$_2$As$_2$ \cite{Chuang2010nematic}, this is the first time that \textit{all} dopants have been imaged in a Ca122 system.

\begin{figure}[tb]
\begin{center}
\includegraphics[width=0.99\columnwidth,clip]{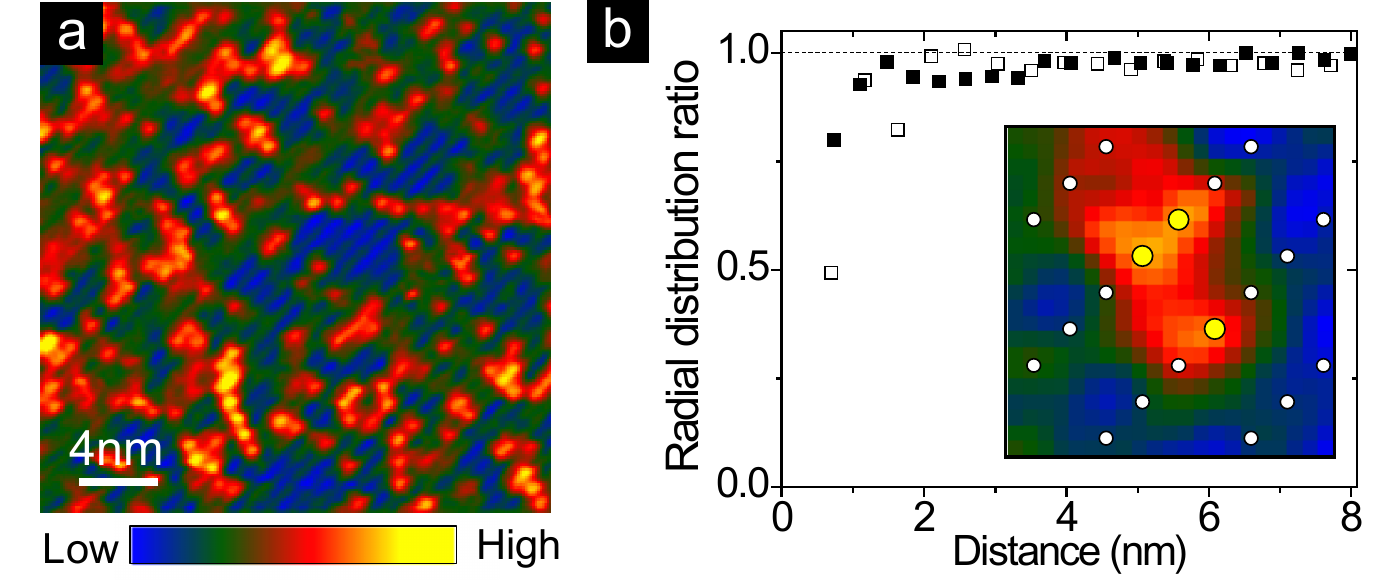}
\caption{(color online) Map of Pr dopants. (a) $dI/dV$ image at +300 mV showing Pr dopants as bright atomic-scale features from the same FOV as Fig.\ 1(b). (b) Radial distribution ratios for two sets of Pr dopants. Full squares represent the distribution of Pr dopants shown in (a), while open squares represent a different dataset used to confirm the conclusions. Inset shows a 2.5 nm $\times$ 2.5 nm region in which surface Ca positions (white dots) and Pr dopants (yellow dots) have been marked, demonstrating our ability to resolve individual Pr dopants even at adjacent Ca sites.
}
\end{center}
\end{figure}

Because we have imaged \textit{all} dopants, we can investigate the possibility of clustering, which was suggested as the origin of the inhomogeneous high-$T_c$ phase \cite{Saha2012structural}. We compute a ``radial distribution ratio'' (RDR) by histogramming all observed Pr-Pr distances within a FOV, then dividing this observed histogram by an average histogram of 1000 simulated random dopant distributions at the same concentration (See Supplementary Information \textbf{S-III} and \textbf{S-IV}). The RDR in Fig.\ 4(b) shows no clustering, and in fact slight repulsion of the Pr dopants at short distances, possibly due to their like charges. The repulsion is not an artifact of poor dopant identification, as illustrated by clear detection of two adjacent Pr dopants in the inset to Fig.\ 4(b). The lack of dopant clustering in \PrCa\ contrasts with the Se dopants in FeTe$_{1-x}$Se$_x$ that are prone to forming patches of $\sim$1 nm size \cite{He2011nanoscale}. This contrast may arise from the $\sim$10\% size mismatch of Se (198 pm) and Te (221 pm) vs.\ the similar sizes of Ca (126 pm) and Pr (126.6 pm) \cite{Shannon1976revised,Song2012electronic}. Our observation of the expected number of Pr dopants, more homogeneously distributed than would be expected for a random distribution, suggests that dopant clustering is unlikely to be responsible for the small volume fraction high-$T_c$ superconducting state. %Note that these 2 FOVs overlap by ~50%, so we should not emphasize the two FOVs.

In conclusion, our STM images of \PrCa\ have addressed its surface structure and dopant distribution, with bearing on its high-$T_c$ volume fraction. First, we used LBH mapping to identify the $2\times 1$ surface as a half-Ca termination, and the $1\times 1$ surface as an As termination. This LBH mapping method could be used to resolve debated cleaved surface terminations in a wide variety of materials, such as other Fe-SCs \cite{Grothe2012LiFeAs} or heavy fermion materials \cite{Schmidt2010URu2Si2,Aynajian2012}. Second, we demonstrated by direct imaging that the Pr dopants responsible for superconductivity do not cluster, and in fact show a slight repulsion at very short length scales. Our findings suggest that Pr inhomogeneity is unlikely to be the source of the high-$T_c$ volume fraction, in contrast to previous speculation \cite{Saha2012structural}.

%Finally, due to the universal difficulty in observing bulk superconductivity on the surface of the Ca122 family, we suggest that Re-doping of Ba122 \cite{Katase2012identical} or Sr122 \cite{Muraba2010high} (perhaps with ions smaller than La (130 pm), such as Nd (124.9 pm) or smaller, to promote the possibility of inducing the structural collapse within those materials) should be carried out and imaged with STM to get to the root of the small volume fraction high-$T_c$, and address fundamental questions about electron vs.\ hole doping in Fe-SCs. % We also propose the alternate way of imaging dopants in materials where elemental work functions are more different than /PrCa?? % SAVE THIS STATEMENT FOR THE CONCLUSION OF THE NEXT PAPER
\begin{acknowledgments}

We thank J.P. Paglione, S. Saha, and E.W. Hudson for helpful conversations. This work was supported by the Air Force Office of Scientific Research under grant FA9550-06-1-0531, and the U.S. National Science Foundation under grant DMR-0847433. D. H. acknowledges support from an NSERC PGS-D fellowship. C. L. S. was supported by the Golub Fellowship at Harvard University.

\end{acknowledgments}

\newpage
\linespread{1.1}
\setlength{\parskip}{2ex}

\renewcommand{\thesection}{S~\Roman{section}}
\renewcommand{\theequation}{S\arabic{equation}}
\renewcommand{\thetable}{S\arabic{table}}
\setcounter{section}{0}

% Renew Figure Counter
\renewcommand{\thefigure}{S\arabic{figure}}
\renewcommand\theHfigure{S\arabic{figure}}
\setcounter{figure}{0}

\onecolumngrid
\begin{center}
\textsc{\textbf{\Large Supplementary Information}}
\end{center}
\vspace{6ex}
\twocolumngrid

\section{S-I.\ Resistive measurements}

\begin{figure}[h!]
\begin{center}
\includegraphics[width=1.0\columnwidth]{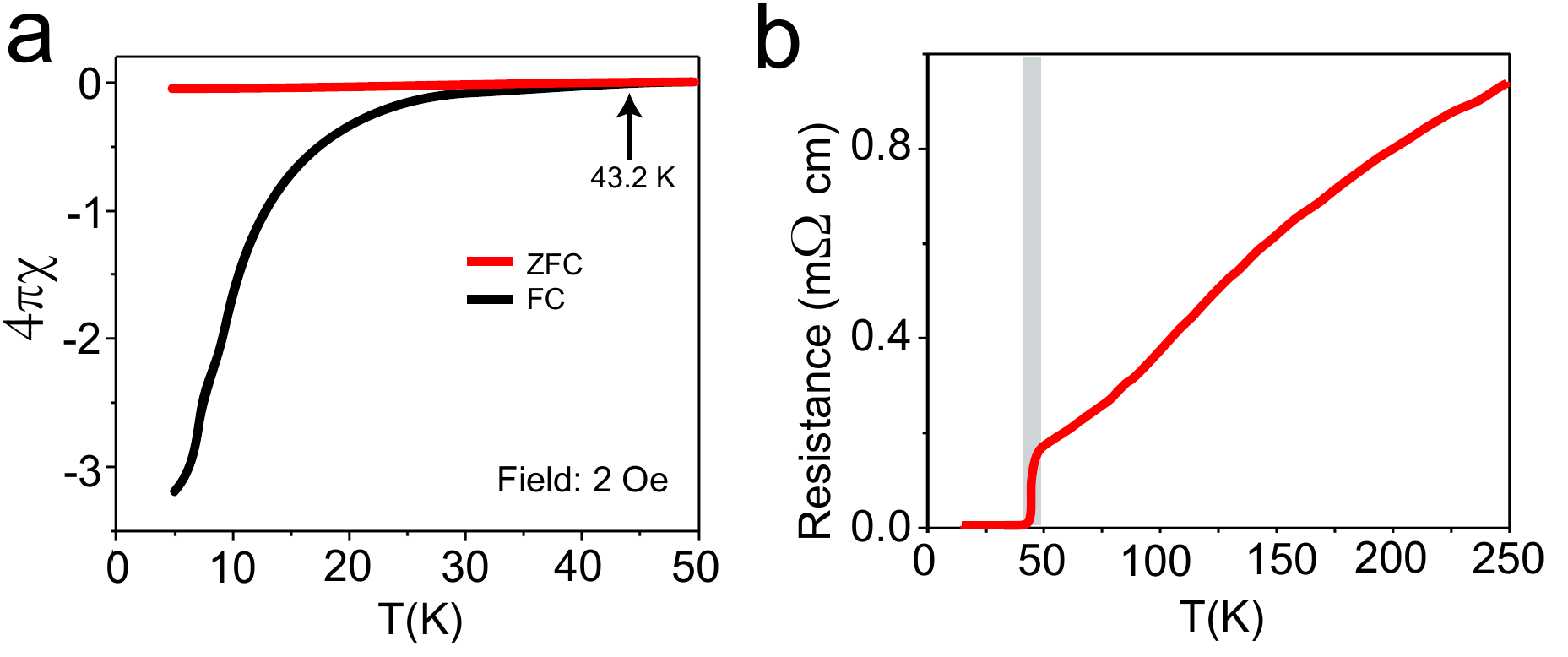}
\caption{\label{fig:Transport} (a) Magnetic susceptibility and (b) resistive transition of the \dPrCa\ sample studied. The gray region denotes the superconducting transition width of 10 K.
}
\end{center}
\end{figure}

The superconducting transition temperature $T_c$=43.2 K in our \dPrCa\ sample is determined by the temperature at which field-cooled magnetic susceptibility deviates from that acquired at zero magnetic field (Fig.~\ref{fig:Transport}(a)). The transition width from the onset of the resistive transition ($\sim$48 K) to zero resistivity ($\sim$38 K), is $\sim$10 K (Fig.~\ref{fig:Transport}(b)). All the measurements presented in this paper were acquired well below 38 K.

\begin{figure}[h!]
\begin{center}
\includegraphics[width=0.83\columnwidth]{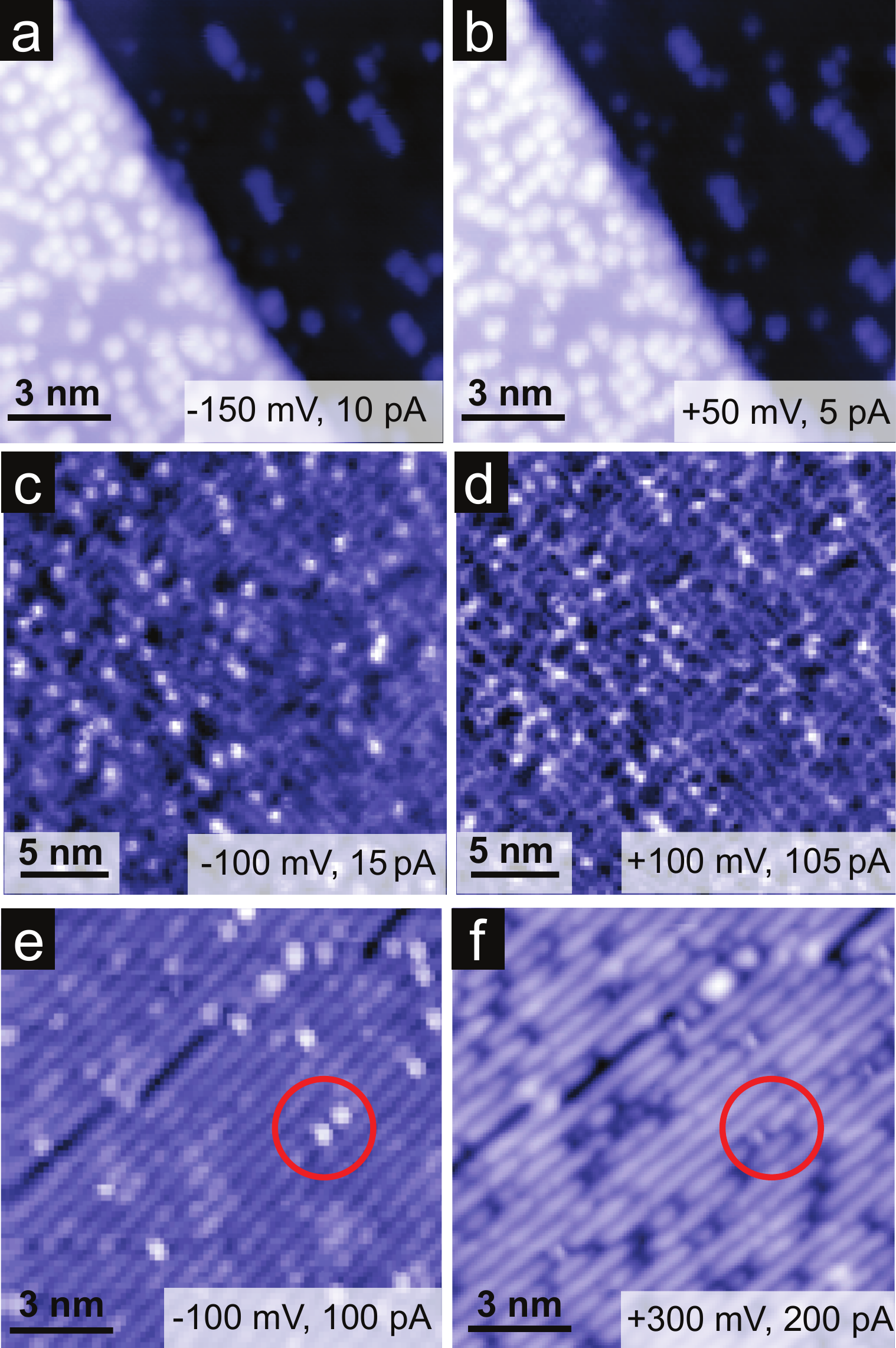}
\caption{\label{fig:BiasDependence}Setpoint dependence of \PrCa\ topographs, used to distinguish between atomic defects lying on top of the surface plane vs. within (or beneath) the surface plane. (a-b) Topographs of region from main text Fig.\ 2(a) using -150 mV and +50 mV sample bias setpoints, respectively. The atomic-sized bright features appear almost identical between the two, suggesting that they are geometric features arising from adatoms lying \textit{on top of} the surface plane. (c-d) Topographs of region from main text Fig.\ 3(a) using -100 mV and +100 mV sample bias setpoints, respectively. The atomic-sized bright features appear in a similar manner between the two, suggesting that they are geometric features arising from adatoms lying \textit{on top of} the surface plane. (e-f) Topographs of region from main text Fig.\ 2(b) using -100 mV and +300 mV sample bias setpoints, respectively. The atomic-sized bright features in (e) disappear at the positive bias used in (f), suggesting that they are electronic features arising from single-atom defects \textit{within} (or beneath) the surface plane. Both (e) and (f) were acquired using the same STM tip.
}
\end{center}
\end{figure}

\section{S-II.\ Setup dependence of impurities in STM topographs}

Not all bright, atomic-scale features seen in STM topographs are adatoms protruding from the surface. Since STM topographs represent a combination of effects due to geometric corrugations and electronic density of states \cite{Tersoff1985}, bright features in topographs might be of electronic origin, coming from surface or sub-surface impurities (which would have no geometric effect on the LBH measurement).

In order to demonstrate this point, Figs.~\ref{fig:BiasDependence}(a-d) show bias-independent atomic-scale topographic features, while Figs.~\ref{fig:BiasDependence}(e,f) show topographic features that vary with the bias setpoint. Thus, we can conclude that bright features in Fig.\ 2(a) and Fig.\ 3(a) of the main text (corresponding to Figs.~\ref{fig:BiasDependence}(a-d)) do represent adatoms, and would therefore have a geometric effect on the LBH. In contrast, bright features in Fig.\ 2(b) (corresponding to Figs.~\ref{fig:BiasDependence}(e,f)) might be resonances due to in-plane or sub-surface impurities, and would therefore not have a geometric effect on the LBH.

\section{S-III.\ Dopant locator algorithm}

\begin{figure*}
\begin{center}
\includegraphics[width=1.75\columnwidth]{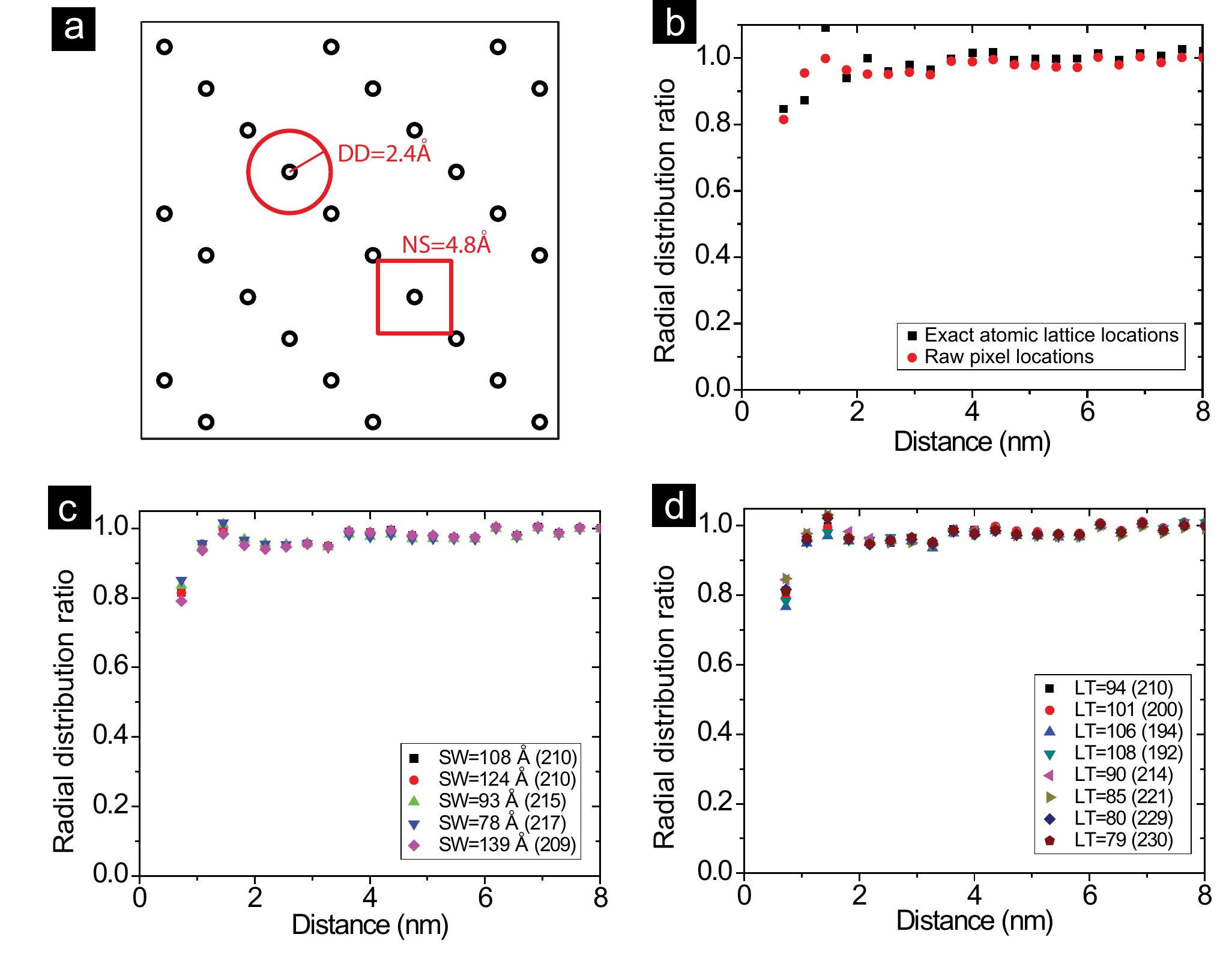}
\caption{\label{fig:DopantLocatorAlg}Robustness of the algorithm used to locate the Pr dopants. (a) Schematic of a perfect 2 x 1 reconstruction, where the small black circles represent the locations of surface Ca/Pr atoms. The red circle and square denote the sizes of DD and NS parameters respectively. (b) RDR computed from the raw pixel positions of the $dI/dV$ local maxima corresponding to Pr dopants (red squares). RDR computed using ``corrected'' Pr dopant positions (black squares). The ``corrected'' positions are calculated by taking the raw integer pixel positions obtained by the algorithm and ``moving'' them to the nearest ideal 2 x 1 lattice site, so that the new position is in fractional units of pixels. In other words, the raw pixel $LM$ position was moved to the nearest exact atom position within a perfect 2 x 1 lattice. The impacts of varying (c) $SW$, and (d) $LT$ parameters on the observed repulsion of Pr dopants. The numbers in parentheses show the number of Pr atoms found by the algorithm for each parameter set. $LT$ is given in arbitrary units.
}
\end{center}
\end{figure*}

The dopant locator algorithm consists of four user-tunable parameters: neighborhood size ($NS$), duplicate distance ($DD$), smoothing window ($SW$), and local threshold ($LT$). For each pixel $(x,y)$ in the field of view (FOV), the algorithm finds the position of the local maximum ($LM$) within a square of side $NS$ centered at $(x,y)$. From the set of $LM$s $(x_{i},y_{i})$, the algorithm eliminates those within $DD$ of a brighter $LM$, and those whose relative brightness compared to the regional average, defined as the average value within a square of side 2$SW$+1 centered at $(x_{i},y_{i})$, is less than $LT$. To identify the dopants in Fig.\ 3(a) of the main paper, we used $DD$=2.4 \AA\ (\textit{cf}.\ Ca-Ca nearest-neighbor distance 4 \AA), $NS$=4.8 \AA, and $SW$=109 \AA. Qualitative conclusions about the absence of clustering and short length scale repulsion of Pr dopants were unaffected by small errors in locating the centers of dopants (Fig.~\ref{fig:DopantLocatorAlg}(b)), and were robust up to 28\% variation of $SW$ (Fig.~\ref{fig:DopantLocatorAlg}(c)) and 15\% variation of $LT$ (Fig.~\ref{fig:DopantLocatorAlg}(d)).

\begin{figure*}
\begin{center}
\includegraphics[width=1.5\columnwidth]{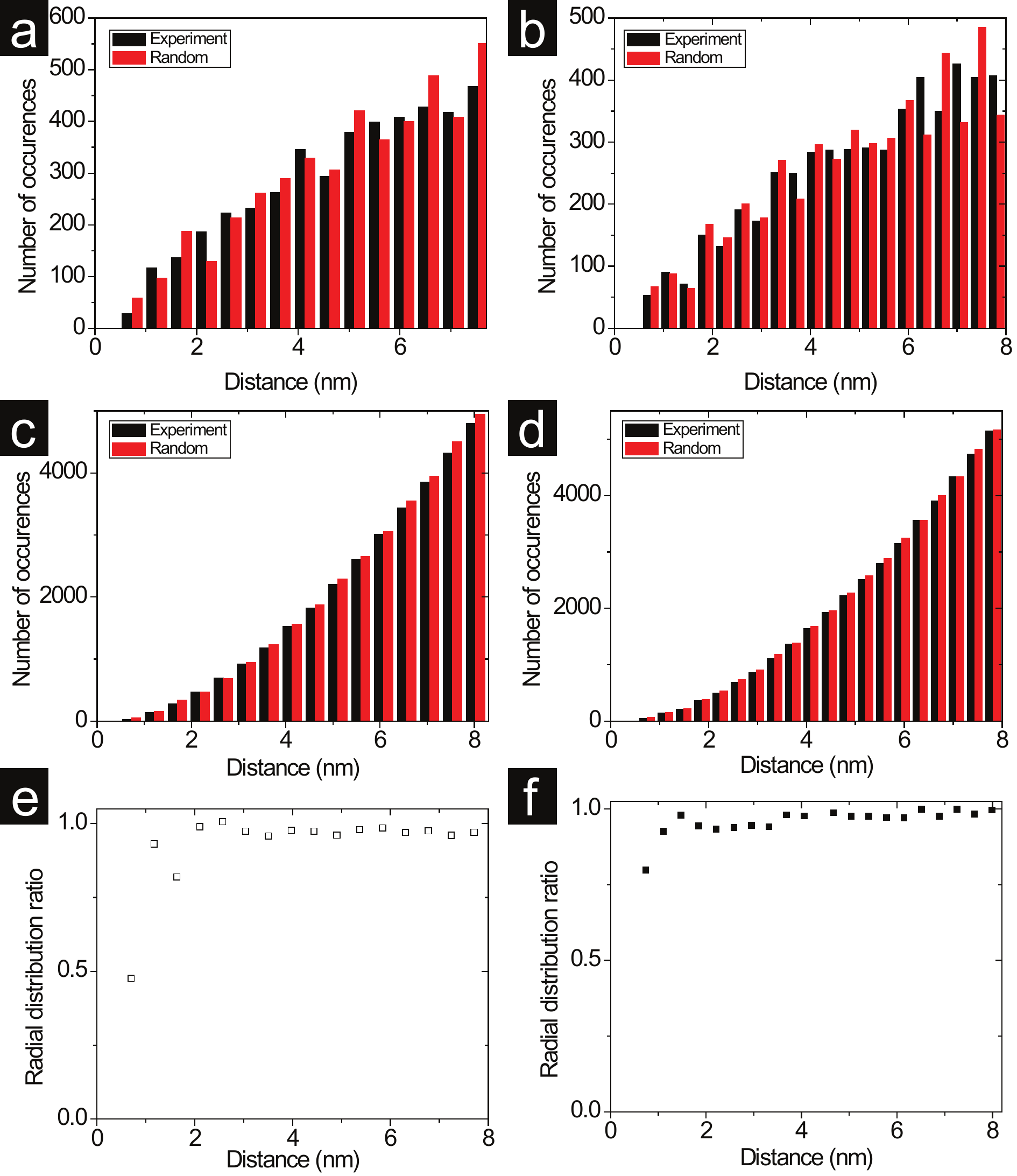}
\caption{\label{fig:Histograms} Histograms of pairwise distances and radial distribution ratios (RDRs). (a,b) Histograms of pairwise distances for completely random (red) and experimentally observed (black) distributions. (c,d) Cumulative histograms of pairwise distances for completely random (red) and experimentally observed (black) distributions. (e,f) RDRs obtained by dividing the black by the red histograms in (c,d) respectively. Left (a,c,e) and right (b,d,f) columns represent two different data sets.
}
\end{center}
\end{figure*}

\section{S-IV.\ Cumulative histograms of pairwise distances}

We visualize the distribution of Pr dopants by plotting non-cumulative (Figs.~\ref{fig:Histograms}(a,b)) and cumulative (Figs.~\ref{fig:Histograms}(c,d)) histograms of pairwise distances for the experimental and randomly generated sets of dopants. The corresponding divided traces of histograms in Figs.~\ref{fig:Histograms}(c,d) are shown in Figs.~\ref{fig:Histograms}(e,f) (same two traces shown in Fig.\ 4(b) of the main text). The random traces in Figs.~\ref{fig:Histograms}(a,b) are generated using an average of N=1000 dopant configurations. We checked that $N$=1000 is sufficiently large, by showing that the result changes by less than 1\% with the use of $N$=5000.


\begin{thebibliography}{39}
\bibitem [{tip()}]{tip}%
  \BibitemOpen
  \href@noop {} {}\bibinfo {note} {Our LBH measurements using PtIr tips on
  polycrystalline Au foils typically yield effective
  $\varphi_{\mathrm{tip}}$=4-5 eV, somewhat less than the elemental work
  functions of the constituants. This observation is also typical in the
  literature, e.g. in Ref.\ 20.}\BibitemShut {Stop}%
\end{thebibliography}
\end{document}